# Renewal-anomalous-heterogeneous files

Ophir Flomenbom

*Flomenbom-BPS, 19 Louis Marshal St., Tel Aviv, Israel 62668*

**Abstract** – Renewal-anomalous-heterogeneous files are solved. A simple file is made of Brownian hard spheres that diffuse stochastically in an effective 1D channel. Generally, Brownian files are heterogeneous: the spheres' diffusion coefficients are distributed and the initial spheres' density is non-uniform. In renewal-anomalous files, the distribution of waiting times for individual jumps is exponential as in Brownian files, yet obeys: $\psi_\alpha(t) \sim t^{-1-\alpha}$, $0 < \alpha < 1$. The file is renewal as all the particles attempt to jump at the same time. It is shown that the mean square displacement (MSD) in a renewal-anomalous-heterogeneous file, $<r^2>$, obeys, $<r^2> \sim <r^2>_{nrml}^\alpha$, where $<r^2>_{nrml}$ is the MSD in the corresponding Brownian file. This scaling is an outcome of an exact relation (derived here) connecting probability density functions of Brownian files and renewal-anomalous files. It is also shown that non-renewal-anomalous files are slower than the corresponding renewal ones.

## 1. Introduction

The stochastic motion of particles determines the behavior of many processes in nature, e.g. [1-4]. An important process in the studies of stochastic dynamics is the diffusion of $N$ ($N \to \infty$) Brownian hard spheres in a quasi-one-dimensional channel of length $L$ ($L \to \infty$) [4-41]. This system is termed file dynamics (sometimes called, single file dynamics). In a basic Brownian file, the spheres have the same diffusion coefficient $D$. The mean particles' density, $\rho$, is fixed: $\rho=\rho_0=N/L$; namely, a constant average microscopic distance among adjacent hard spheres, $\Delta$ (=$L/N$), characterizes the system.

The applicability of file dynamics is vast. It models many microscopic processes that occur in nature and in applications. Several examples include [1, 30-40]: *(a)* Diffusion within biological and synthetic pores and in porous materials of water, ions, proteins, and organic molecules [1, 30]. (*b*) Diffusion along 1D objects, such as the motion of motor-proteins along filaments [1]. (*c*) Conductance of electrons in nano-wires [40]. *(d)* File dynamics has also been related to the dynamics of a monomer in a polymer [29, 34].

In a *basic* Brownian file, the mean square displacement (MSD) $<r^2>$ of a tagged particle in the file scales as: $<r^2> \approx (Dt)^{1/2}/\rho_0$, where its corresponding PDF is a Gaussian (in position). These are the most well-known statistical properties of basic Brownian files. Still, in realistic systems, one, or several, of the conditions defining the *basic* Brownian file may break down in a way that leads to different dynamical behaviors. For example, in some channels, the particles may bypass each other with a constant probability upon collisions [19-22], and this may lead to an enhanced diffusion at long times. Yet, when the particles interact with the channel, a slower diffusion is observed [15]. When the initial particles' density law is non uniform [23]:

$$\rho(l) \sim \rho_0 (l\rho_0)^{-a}, \qquad 0 \leq a \leq 1, \qquad (1)$$

enhanced diffusion is seen. The file's middle particle's MSD is expressed in terms of the free particle's MSD, $<r^2>_{free}$ [23]:

$$<r^2> \sim \rho_0^{a-1} <r^2>_{free}^{(1+a)/2}. \qquad (2)$$

$\rho(l)$ in Eq. (1) is the initial density of the file; namely, the particles are positioned at, $x_{0,j}$=sign($j$)$\Delta |j|^{1/(1-a)}$, for $|j| \leq M$, where, $N = 2M + 1$. Equation (1) means that at the initial stage of the process, the number of particles $n$ as a function of the length $l$ from the origin obeys, $n \sim (l\rho_0)^{1-a}$. Physically, this means that the distance between particles increases as we look on particles positioned at larger and larger distances from the origin. As Eq. (2) is for the file's middle particle, it reflects the expansion process from the denser region in the middle of the file to the dilute region in the periphery. Note that the density law in Eq. (1) is a simple way to study the effect of a non-uniform particles' density on the dynamics in the file. We expect that most real files naturally obey the condition of a non-constant density, where Eq. (2) reflects files that are very inhomogeneous in their density.

Equation (2) was derived analytically in Ref. [23] for a Brownian file, and was extended for *any* renewal file using scaling law analysis. Here, a renewal file is a file in which all the particles attempt to jump at same time, and the time for each attempt is an independent random variable. (This condition is automatically fulfilled in a Brownian file, but not in general.) We use the term a renewal process in accordance with this term in probability theory [42].

Equation (2) is pretty general, yet limited to the other conditions of a basic file; for example, Eq. (2) holds in a file that all its particles have the same diffusion coefficient. Still, in many files the diffusing particles are not identical. We have recently studied Brownian files with a distribution in the diffusion coefficients. We termed such files heterogeneous files. Here, each particle's diffusion coefficient is taken randomly and independently from the PDF,

$$W(D) \sim (1-\gamma)\Lambda^{-1}(D/\Lambda)^{-\gamma} \qquad , \qquad 0 \leq \gamma < 1. \qquad (3)$$

In Eq. (3), $\Lambda$ is the fastest possible diffusion coefficient in the file. The initial conditions are still distributed according to Eq. (2). Using analytical calculations, we have showed that the MSD for the tagged particle in a heterogeneous-Brownian files reads [28],

$$\rho_0^2 < r^2 >_{nrml} \sim (\rho_0^2 \Lambda t)^\mu \qquad ; \qquad \mu = \frac{1-\gamma}{2/(1+a)-\gamma}, \qquad (4)$$

where the corresponding PDF is a Gaussian. (In this Letter, $\mu$ always represents the scaling power of the MSD of *Brownian* files.) Importantly enough, when setting $\gamma = 0$ in Eq. (4), one sees that the scaling of the MSD of the heterogeneous file is the same as that of a basic Brownian file, Eq. (2). Thus, Eq. (4) finds the condition when $W(D)$ affects the dynamics. Namely, when $W(D)$ is flat, the fact that the file is heterogeneous is not observed when calculating the MSD; only when $W(D)$ gives a significant weight to small diffusion coefficients, the heterogeneity affects the dynamics.

In this Letter, we consider an important extension of heterogeneous-Brownian files: renewal-anomalous-heterogeneous files. In a renewal-anomalous file, all the particles attempt to jump *at same time*, after residing in their positions for exactly the same period of time. This random period is drawn each time independently from a waiting time (WT-) PDF of the form: $\psi_\alpha(t) \sim k(kt)^{-1-\alpha}$, $0 < \alpha < 1$, where $k$ is a parameter. This is a renewal file: there is only one clock in the system, and each trajectory of a given particle is a renewal process in the sense that the waiting periods are independent random variables. The particles' diffusion coefficients are still drawn from $W(D)$, Eq. (3), and the particles' density $\rho(l)$ is not fixed, and obeys Eq. (1).

The physical model of renewal-anomalous files can relate pretty naturally to the dynamics of particles in fluctuating pores. Possible realizations of such files include: pores under on-off fields or under temperature changes (say, controlled externally), sensing devices (as was suggested for zeolites, e.g. [44]) under on-off fields, and channels as sequencing devices (e.g. [45]) under on-off fields. Let us also present such a possible system in detail: it contains a channel that occupies one of two possible states, a state that enables motion, and a state that doesn't. When the channel occupies the later state, the particles can easily bind to the channel. The dynamics of the complex process consist of the following stages: * The particles diffuse in a channel; the channel is in a mode that enables motion. * At random times, the channel switches to a mode that facilitates the binding of particles to the channel. In this mode of the channel, all the particles bind to the channel very quickly. * The particles disassociate from the channel at random times, yet simultaneously, as this depends on the time that the channel changes its mode to a mode that enables motion. These three stages form the model of the diffusion in a fluctuating channel. Now, in some cases, the stochastic binding times may indeed be distributed according to a PDF of the form of $\psi_\alpha(t)$. Recall that we have rationalize the synchronized disassociation of the particles from the pore as a result of large scale fluctuations in the channel shape; yet, the reason for a power-law WT-PDF may be attributed to the interactions of the channel with a heterogeneous medium. It is known that the influence of a heterogeneous medium on a diffusion object can lead to a power-law WT-PDF of the form of $\psi_\alpha(t)$ for the diffusing object, e.g. [43]. Now, it is well-known that bio-channels in physiological conditions change their structure constantly [1] and are in contact with a membrane that is heterogeneous in composition. Thus, renewal-anomalous files may indeed serve as a promising choice for modeling biological pores in not so few cases.

With the physical picture and several possible real-life realizations in mind, we focus our attention on solving renewal-anomalous files. We prove here analytically that the MSD for renewal-anomalous-heterogeneous files scales as the MSD of the corresponding Brownian files raised to the power of $\alpha$:

$$< r^2 > \sim < r^2 >_{nrml}^\alpha, \qquad (5)$$

where $< r^2 >_{nrml}$ appears in Eq. (4). Equation (5) is an outcome of a general relation connecting PDFs of Brownian files and renewal-anomalous files; this relation, Eq. (10), is proved here for the first time. Equation (5) generalizes our previous results in Eq. (2), where the file's MSD is related to free particle's MSD of the same dynamics. Now, we also show here that Eq. (5) can be obtained using scaling law analysis, where such an analysis further explains the relation among Brownian files and renewal-anomalous files. Finally, we show that when the file is non-renewal and anomalous, the relation in Eq. (5) does not hold. Non-renewal anomalous files are much slower than their renewal counterparts.

**2. Renewal-anomalous files**

In this part, we first prove Eq. (5) analytically. We then come to the same results using scaling-law analysis. Finally, we present results from extensive numerical simulations of renewal-anomalous-heterogeneous files, which also support the relation in Eq. (5). A short discussion on non-renewal-anomalous files is also presented. Yet, we start the discussion with a clear definition of the physical model, using a simulation scheme [the mathematical definition of renewal-anomalous files is presented in Eq. (9)].

*2.1. The definition of a renewal-anomalous file* A very clear way to define the physical model of renewal-anomalous files uses a simulation scheme. We suggest the following scheme: * a random waiting time $\tau$ is drawn independently from $\psi_\alpha(t)$, * all the particles in the file stand still for this random period, * after residing in their positions for a time period $\tau$, all the particles try to jump according to the standard rules of the file. * This procedure is carried on over and over and over again.

*2.2. Analytical calculations* To analyze the dynamics of renewal-anomalous-heterogeneous files, we write the equation of motion for the *N*-particle PDF of the file. This equation is obtained from the equation of motion for a Brownian file when applying a simple convolution on it. So, we first write the equation of motion for a Brownian-heterogeneous file; this reads,

$$\partial_t P_{nrml}(x,t \mid x_0) = \sum_{j=-M}^{M} D_j \partial_{x_j} \partial_{x_j} P_{nrml}(x,t \mid x_0) \equiv L_\mu P_{nrml}(x,t \mid x_0). \qquad (6)$$

Here, $P_{nrml}(x,t \mid x_0)$ is the PDF that the file's particles are located at positions $x$, $x = (x_{-M}, x_{-M-1}, \ldots, x_M)$, at time $t$ starting from an initial condition $x_0$ at time, $t_0 = 0$. Equation (6) is solved with the appropriate boundary conditions, which reflect the hard-sphere nature of the process:

$$(D_j \partial_{x_j} P_{nrml}(x,t \mid x_0))_{x_j=x_{j+1}} = (D_{j+1} \partial_{x_{j+1}} P_{nrml}(x,t \mid x_0))_{x_{j+1}=x_j} \; ; \quad j = 1, \ldots, N-1, \qquad (7)$$

and with the appropriate *initial* condition:

$$P_{nrml}(x, t \to 0 \mid x_0) = \prod_{j=-M}^{M} \delta(x_j - x_{0,j}) \; ; \quad x_{0,j} = \Delta \frac{j}{|j|}(\Delta |j|)^{1/(1-a)}. \qquad (8)$$

The PDF for a renewal-anomalous-heterogeneous file is obtained from Eqs. (6)-(7) when convoluting them with a kernel $k_\alpha(t)$. In particular, the equation of motion reads:

$$\partial_t P(x,t \mid x_0) = \sum_{j=-M}^{M} D_j \partial_{x_j} \partial_{x_j} \int_0^t k_\alpha(t-u) P(x,u \mid x_0) du. \qquad (9)$$

The kernel $k_\alpha(t)$ in Eq. (9) is related to the WT-PDF $\psi_\alpha(t)$ of the same renewal dynamics; this relation is made in Laplace space (e.g. [3]):

$$\bar{k}_\alpha(s) = \frac{s\bar{\psi}_\alpha(s)}{1-\bar{\psi}_\alpha(s)},$$

where the Laplace transform of a function $f(t)$ reads, $\bar{f}(s) = \int_0^\infty f(t) e^{-st} dt$.

We note that Eq. (9), for a uniform file ($D_j = D$), was introduced in Ref. [27]. Yet, we emphasize here that Eq. (9) holds *only* for renewal-anomalous files; that is, for files in which all the particles attempt to jump at the same time. This is the reason that Eq. (9) has a form of a simple convolution. A *non-renewal*-anomalous file, in which each particle has its own jumping-clock, should have a different equation of motion, leading to a different dynamical behavior than that of renewal-anomalous files. We shortly discuss non-renewal-anomalous files in the next paragraph, and show that such files are slower than their renewal counterparts.

Now, we continue with the analysis of renewal-anomalous files, Eq. (9), and write the PDF for the system in terms of the PDF that solves the un-convoluted equation, Eq. (6). The relation is made in Laplace space:

$$\bar{P}(x, s \mid x_0) = \frac{1}{\bar{k}_\alpha(s)} \bar{P}_{nrml}(x, s/\bar{k}_\alpha(s) \mid x_0). \tag{10}$$

Equation (10) is a central result of this Letter: it relates renewal-anomalous files and the corresponding Brownian files. To prove Eq. (10), we formally solve Eq. (9) in Laplace space:

$$\bar{P}(x, s \mid x_0) = \left(s + \bar{k}_\alpha(s) L_\mu\right)^{-1} P_{nrml}(x, 0 \mid x_0) = \frac{1}{\bar{k}_\alpha(s)} \left(\frac{s}{\bar{k}_\alpha(s)} + L_\mu\right)^{-1} P_{nrml}(x, 0 \mid x_0), \tag{11}$$

and notice that Eq. (6) has a Laplace space solution of the form,

$$\bar{P}_{nrml}(x, s \mid x_0) = \left(s + L_\mu\right)^{-1} P_{nrml}(x, 0 \mid x_0).$$

When using this equation in rewriting the last expression in Eq. (11), we obtain Eq. (10).

From Eq. (10), it is straightforward to relate the MSD of normal heterogeneous files and renewal-anomalous heterogeneous files,

$$<\bar{r}^2(s)> = \frac{1}{\bar{k}_\alpha(s)} <\bar{r}^2(s/\bar{k}_\alpha(s))>_{nrml}. \tag{12}$$

Now, from Eq. (4) we have, $<r^2(t)>_{nrml} \sim t^\mu$, and in Laplace space, $<\bar{r}^2(s)>_{nrml} \sim s^{-1-\mu}$, and so Eq. (12) gives,

$$<\bar{r}^2(s)> = \frac{1}{\bar{k}_\alpha(s)}(s/\bar{k}_\alpha(s))^{-1-\mu}.$$

Using the asymptotic form of $\bar{\psi}_\alpha(s)$ (small $s$), $\bar{\psi}_\alpha(s) \sim 1 - (sT)^\alpha$, we find the kernel in Laplace space, $\bar{k}_\alpha(s) \sim (sT)^{1-\alpha}$, and $<\bar{r}^2(s)>$ follows,

$$<\bar{r}^2(s)> \sim \frac{s^{-\alpha(1+\mu)}}{s^{1-\alpha}} = s^{-1-\alpha\mu}.$$

This equation reads in time-space,

$$<r^2(t)> \sim t^{\alpha\mu}. \tag{13}$$

The above expression for the MSD of a renewal-anomalous file has a very appealing consequence: one can use the results of a file with normal dynamics raised to the power of $\alpha$ for obtaining the results for the corresponding renewal-anomalous file,

$$<r^2(t)> \sim <r^2(t)>_{nrml}^\alpha. \tag{14}$$

Equations (13)-(14) are among the main results of this paper. Equations (13)-(14) originated from a general relation connecting PDFs of renewal-anomalous files and Brownian files, Eq. (10). In the next paragraph, the same results are derived using scaling law analysis. This unravels another interesting relation connecting renewal-anomalous files and Brownian files.

***2.3. Scaling law analysis*** In this paragraph, we derive a similar relation to Eq. (13) in a way that gives additional insights into the behavior of renewal-anomalous files. First, we realize that a Brownian file is a renewal file in which all the particles attempt a jump every time step $dt$. This is simply seen when simulating the discrete-time-version of the equation of motion, Eq. (5). A consequence of this property is that the average number of attempts to jump as a function of the time, $<J(t)>$, scales, for any particle in the heterogeneous-Brownian-file, as, $<J(t)> \sim t$. This is found from the general relation for $<\bar{J}(s)>$ for a renewal process with a WT-PDF $\psi(t)$, e.g. [4], $<\bar{J}(s)> = \frac{\bar{\psi}(s)}{s(1-\bar{\psi}(s))}$, when using the fact that $\psi(t)$ is exponential for a Brownian file. Now, when using this relation for $\psi_\alpha(t)$, we find that for a renewal-anomalous file, $<J(t)> \sim t^\alpha$. The above is used in the following way. First, we recall that for any renewal dynamics, $<r^2(t)>_{free} \sim <J(t)>$, e.g. [4], and use this in Eq. (2) for writing, $<r^2(t)> \sim <J(t)>^{(1+a)/2}$. The next step uses the above also in Eq. (5),

$$<r^2(t)> \sim <J(t)>^\mu.$$

Taking this relation to hold for any renewal process, it is the same as Eq. (13).

There is another way for using $<J(t)>$ in relating a Brownian file with a renewal-anomalous file. Here, we take the trajectory that changes its value every time step $dt$, and stretch each time step $dt$ to a random period drawn from the WT-PDF, $\psi_\alpha(t)$. Clearly, this manipulation takes a trajectory with, $<J(t)> \sim t$, and makes it a trajectory with, $<J(t)> \sim t^\alpha$. This suggests using the transformation $t \to t^\alpha$ in $<r^2(t)>_{nrml}$ for obtaining $<r^2(t)>$; namely:

$$<r^2(t)> \sim <r^2(t^\alpha)>_{nrml}. \tag{15}$$

We will use the above manipulation in the numerical calculations presented in the next paragraph.

***2.4. Numerical simulations*** Based on the above scheme for simulating renewal-anomalous files, we present in this paragraph the results from extensive simulations. First, Fig. 1 shows a pair of trajectories: the left trajectory is obtained from a simulation of Eq. (5) with $a = 1/3$ and $\gamma = 0$, and the right trajectory is obtained from the left trajectory when applying the manipulation described in paragraph ***2.3.***, with $\alpha = 1/3$. Note that the right trajectory is stretched one hundred thousand times due to the timescale manipulation, forming the whole left trajectory.

Now, for each renewal-anomalous trajectory, such as the right trajectory in Fig.1, we calculate the MSD. Note that the calculations of the MSD from renewal-anomalous trajectories demand taking into account the fact that the original time vector has random increments. The most efficient way to calculate the MSD for such a form of the time vector creates, for each value of $t$ in $<r^2(t)>$, a trajectory that is monitored in time interval of length $t$, and from this trajectory calculates the value of $<r^2(t)>$. The results for the MSD for renewal-anomalous files are shown in the four panels of Fig. 2. Each panel has a constant value of $a$ and $\gamma$, taken from the following values: $a = 0, \frac{1}{3}$, and, $\gamma = 0, \frac{1}{3}$, and shows two curves of the MSD for the various values of $\alpha$, $\alpha = \frac{1}{3}, \frac{2}{3}$, where $\alpha = \frac{1}{3}$ for the lower curve. Also shown are the analytical curves from Eq. (13). The curves from the simulations coincide nicely with the analytical curves. Note that the results for the MSD with $\alpha = \frac{1}{3}$ span twelve orders of magnitude.

We also present here preliminary results from simulations of non-renewal-anomalous files. (A comprehensive study of such files is the subject of our forthcoming publication in this field.) In the left panel of Fig. 3, a trajectory from a simulation of a non-renewal-anomalous file is shown in blue (upper curve in this panel). For making a comparison explicit, a trajectory from a *corresponding* renewal-anomalous file is also plotted (lower curve, black). The two trajectories are plotted as a function of the event index. Clearly, this panel shows that only the non-renewal trajectory shows anomalous patterns when plotted as a function of its indices. Importantly, the time vectors of non-renewal-anomalous files and renewal-anomalous files evolve in a similar way; this is shown in the right panel in Fig. 3 that plots these time vectors as a function of their indices. Basically, this figure shows that non-renewal-anomalous files are much slower than their renewal counterparts.

## 3. Concluding remarks

In this Letter, the dynamics of renewal-anomalous-heterogeneous files were considered. The heterogeneity is evident in both the initial particles' density $\rho$ at length $l$ from the origin, obeying, $\rho(l) \sim \rho_0 l^{-a}$, $0 \leq a \leq 1$, and in the file's diffusion coefficients, which are distributed according to the PDF, $W(D) \sim D^{-\gamma}$, $0 \leq \gamma < 1$, for small $D$. The underlying dynamics are anomalous: any waiting time is taken independently from the PDF for individual jumps obeys: $\psi_\alpha(t) \sim k^\alpha t^{-1-\alpha}$, $0 < \alpha < 1$. We have showed here that in a renewal-anomalous file, in which all the particles attempt to jump at the same time, the mean square displacement (MSD) of a particle in the file scales as, $<r^2> \sim <r^2>^\alpha_{nrml}$, where $<r^2>_{nrml}$ is the result for the MSD in a corresponding Brownian file. This relation originates from a general relation connecting PDFs of renewal-anomalous files with those of Brownian files. A microscopic explanation for these relations was supplied: a trajectory of any particle in a renewal-anomalous file can be obtained from a trajectory of a corresponding particle in a Brownian file when stretching each time increment in the Brownian trajectory to a random length drawn independently from $\psi_\alpha(t)$. This basically leads to the relation, $<r^2(t)> \sim <r^2(t^\alpha)>_{nrml}$. Finally, it was also shown here that non-renewal-anomalous files are slower than their renewal counterparts. In such files, the particles are anomalous, yet each has its own clock of jumping times, meaning that Eq. (5) does not describe the dynamics, and this indeed has a critical effect on the dynamics. We will elaborate on such files in a forthcoming publication.

**FIGURE CAPTIONS**

**Figure 1** Trajectories from file dynamics. The left trajectory is obtained from a simulation of Eq. (5), with, $D_j \rightarrow D$. The right trajectory is obtained from the left trajectory when applying the time-scale manipulation described in paragraph 2.3., with, $\alpha = 1/3$ and $k = 1$. In the simulation, $N = 501$, $dt = 0.13$, $\Delta = 1$, $D = 1$ and $a = 1/3$.

**Figure 2** The MSD, on a log-log scale, from extensive simulations for the various values of $a$, $\gamma$ and $\alpha$. Each panel has distinct values of $a$ and $\gamma$, written explicitly on the panel, yet the value of $\alpha$ varies, $\alpha = \frac{1}{3}, \frac{2}{3}$. The lower curve in each panel corresponds to $\alpha = 1/3$. The curves from our estimation for the MSD are also presented. The coincidence with the results from the simulations is pretty clear in all cases.

**Figure 3** Trajectories from a non-renewal anomalous file (blue online, upper curve) and renewal-anomalous-file (black online, lower curve), as a function of the event index, are shown on the left panel. For both trajectories, $a = 1/3$, $\gamma = 0$ and $\alpha = \frac{1}{3}$ (the other file's information is as in Fig. 1). Clearly, the non-renewal-anomalous trajectory shows anomalous patterns and not its renewal counterpart. The time vectors of both trajectories are of the same magnitude anywhere (right panel). In this right panel the lower curve corresponds to the time vector of the renewal-anomalous file.

**FIGURE 1**

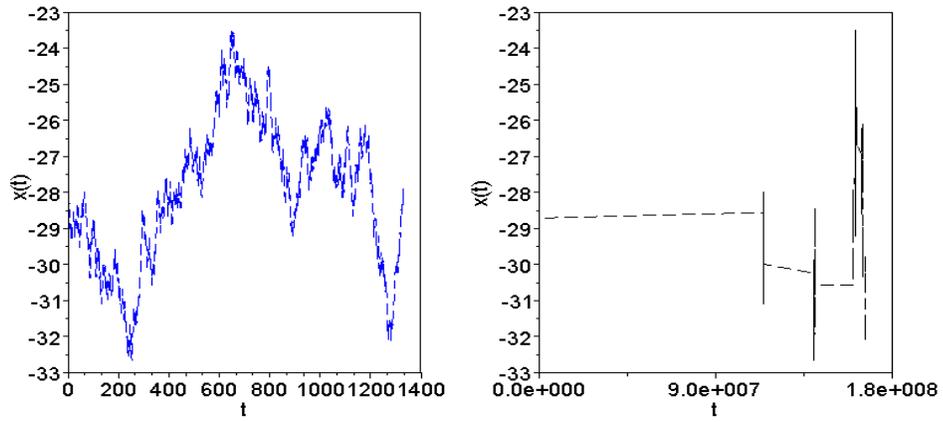

**FIGURE 2**

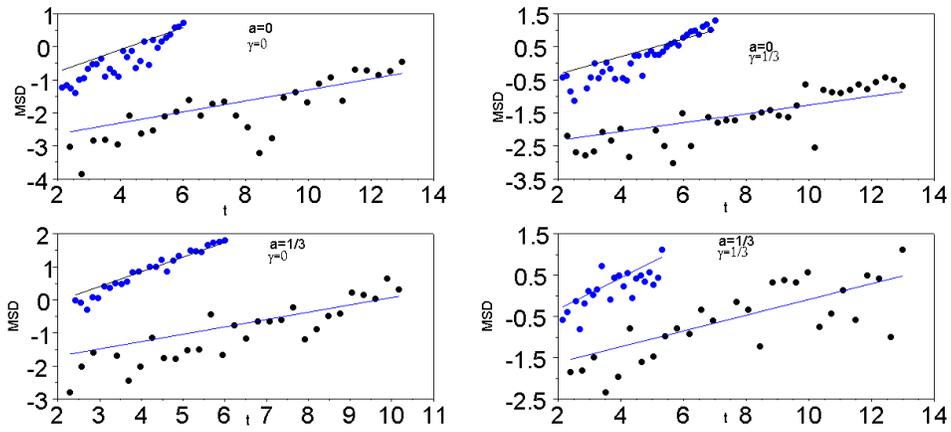

**FIGURE 3**

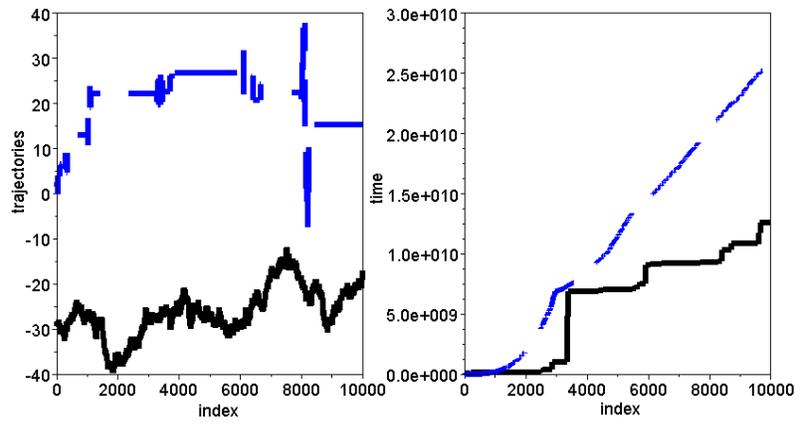